\documentclass[twocolumn,nobibnotes,nofootinbib]{revtex4}

\usepackage{amsmath,amssymb,bm}
\usepackage{epsfig}
\usepackage{graphicx}
\usepackage{pstricks,pst-node,pst-text,pst-3d}
\usepackage{subfigure}
\usepackage{hyperref}

\newcommand{\rr}{\bm{r}}

\def\beq{\begin{equation}}
\def\eeq{\end{equation}}
\begin{document}

\title{Estimating non-linear QCD effects in  ultrahigh energy neutrino events at IceCube }
\author{V.P. Gon\c{c}alves$^{1}$ and D. R. Gratieri$^{1,2}$}
\affiliation{$^{1}$
Instituto de F\'{\i}sica e Matem\'atica,  Universidade
Federal de Pelotas, 
Caixa Postal 354, CEP 96010-900, Pelotas, RS, Brazil}

\affiliation{$^{2}$
Instituto de F\'{\i}sica Gleb Wataghin - UNICAMP, 13083-859, Campinas, SP, Brazil
}

\begin{abstract}

The number of ultrahigh energy events at IceCube is estimated, for the first time, taking into account non-linear QCD effects in the neutrino - hadron cross section. We assume that the  extragalactic neutrino flux is given by $\Phi_{\nu}(E_{\nu})=\Phi_{0}E^{-2}_{\nu}$ and estimate the neutrino - hadron cross section using the dipole approach and a phenomenological model for the dipole - hadron cross section based on non-linear QCD dynamics. We demonstrate that the non-linear prediction is able to describe the current IceCube data and that the magnitude of the non-linear effects is larger than 20 \% for visible energies of order of 2 PeV and increases with the neutrino energy. Our main conclusion is that the non-linear QCD effects are non-negligible and should be taken into account in the analysis of the number of ultrahigh energy events.

\end{abstract}

%
\maketitle

Observations in the last years of the neutrino events with deposited energies ranging from 30 to 2000 TeV by the IceCube Observatory \cite{ice:science,ernie,icedata} has opened a new era in the Neutrino Physics, motivating a lot of studies about the production, composition, propagation and detection of neutrino at ultrahigh energies (For a recent review see, e.g. Ref. \cite{halzen}). One { of} the main ingredients in these studies is the neutrino -  hadron cross section ($\sigma_{\nu h}$), which is probed in its high energy behaviour. In our previous studies \cite{hepp,vicdie} we have estimated the theoretical uncertainty present in the predictions for $\sigma_{\nu h}$ associated to the treatment of the QCD dynamics at high energies and demonstrated that the different predictions can differ by a factor 1.5  for neutrino energies around  $10^{6}$ GeV and increases to 5.5 for   $E_{\nu} = 10^{13}$ GeV. These results motivate the study of the impact of the QCD dynamics in the predictions of the number of events at IceCube, being it the  main aim of this letter. As in Ref. \cite{Soni} the authors have performed a detailed analysis of the Standard Model expectations, taking into account the theoretical uncertainties associated mainly to the parton distribution functions, which are solution of the {\it linear} QCD dynamics described by the Dokshitzer - Gribov - Lipatov - Altarelli - Parisi (DGLAP) evolution equations \cite{dglap}, our main focus will be to complement that study taking into account {\it non-linear} QCD effects in the calculation of the number of ultrahigh energy neutrino events.  As discussed in detail in Refs. \cite{hepp,vicdie},  a transition from the linear DGLAP dynamics  to a new regime where the physical process of recombination of partons becomes important in the parton cascade and the evolution is given by a non-linear evolution equation \cite{bal,cgc,kovchegov} is expected at high energies.  Our goal is to estimate the magnitude of these new dynamical effects in the current energies probed in the IceCube Observatory and present our predictions for higher energies.

Initially lets present  the formalism for the calculation of  the number of neutrino  events in IceCube. Following \cite{ice:science} we define the number of events as
\begin{eqnarray}
dN= T\Omega \sum_{\nu+\bar \nu} N_{eff}(E_{\nu})\sigma_{\nu h}(E_{\nu})\Phi_{\nu}(E_{\nu})~dE_{vis}
\label{eq:dn}
\end{eqnarray}
where $T=998$ days of data-taken, $\Omega=4\pi$ is the solid angle  and $N_{eff}$ is the effective number of scatters in the detector, which can be related with the effective volume $V_{eff}$ through the Avogadro's number $N_{A}$. $V_{eff}$ can be written  in terms of the effective mass $M_{eff}$ and the effective ice density $\rho_{eff}$ as given in Ref. \cite{ice:science}. As the neutrino astrophysical sources are too far from Earth,  all the neutrino flavours are equalized at time they reach the detector due to neutrino oscillations, such that $\nu_{e}:\nu_{\mu}:\nu_{\tau} = 1:1:1$. As in Ref. \cite{Soni} we will assume that the power spectrum  $\phi_{\nu}$ is given by 
\begin{equation}
\Phi_{\nu}(E_{\nu})=\Phi_{0}E^{- l}_{\nu}
\label{eq:flux}
\end{equation} 
with  the overall normalization per flavour being given 
\begin{equation} 
\Phi_{0}=1.2\frac{10^{-8}GeV}{s~cm^{2}sr}~.
\end{equation}
and $l=2$ \cite{ice:science}.
The total cross sections are given by \cite{book}
\begin{eqnarray}
\sigma_{\nu N}^{CC,\,NC} (E_\nu) = \int_{Q^2_{min}}^s dQ^2 \int_{Q^2/s}^{1} dx \frac{1}{x s} 
\frac{\partial^2 \sigma^{CC,\,NC}}{\partial x \partial y}\,\,,
\label{total}
\end{eqnarray}
where $E_{\nu}$ is the neutrino energy, $s = 2 ME_{\nu}$ with $M$ the nucleon mass, $y = Q^2/(xs)$ and $Q^2_{min}$ is the minimum value of $Q^2$ which is introduced in order to stay in the deep inelastic region. In what follows we assume $Q^2_{min} = 1$ GeV$^2$. Our results are almost insensitive to this choice, since the $Q^2$ integral is dominated by values of the order of the electroweak boson mass squared. Moreover, the differential cross section is given by \cite{book}
 \begin{widetext}
\begin{eqnarray} 
\frac{\partial^2 \sigma_{\nu N}^{CC,\,NC}}{\partial x \partial y} = \frac{G_F^2 M E_{\nu}}{\pi} \left(\frac{M_i^2}{M_i^2 + Q^2}\right)^2 \left[\frac{1+(1-y)^2}{2} \, F_2^{CC,\,NC}(x,Q^2) - \frac{y^2}{2}F_L^{CC,\,NC}(x,Q^2) \right. \nonumber\\
\left. + y (1-\frac{y}{2})xF_3^{CC,\,NC}(x,Q^2)\right]\,\,,
\label{difcross}
\end{eqnarray}
 \end{widetext}
where $G_F$ is the Fermi constant and $M_i$ denotes the mass of the charged or neutral gauge boson. 
The calculation of $\sigma_{\nu h}$ involves  integrations over $x$ and $Q^2$, with the integral being dominated by  the interaction with partons of lower $x$ and  $Q^2$ values of the order of the electroweak boson mass squared. 
In the QCD improved parton model the structure functions $F_2,  \,F_L$ and $F_3$ are calculated in terms of quark and gluon distribution functions. In this case the neutrino - hadron cross section for charged current interactions on an isoscalar target is given in terms of the parton distribution functions (See, e.g. Ref. \cite{book}).
However, as discussed in Refs. \cite{hepp,vicdie}, in order to estimate the  non-linear  effects in the QCD dynamics, it is more adequate to  describe  the structure functions considering the color dipole approach,   in which the neutrino - hadron scattering can be viewed as a result of the interaction of a color $q\bar{q}$ dipole which the gauge boson fluctuates \cite{nik}.  In this approach the $F_2^{CC,\,NC}$ structure function is expressed in terms of the transverse and longitudinal structure functions, $F_2^{CC,\,NC}=F_T^{CC,\,NC} + F_L^{CC,\,NC}$ which are given by 
\begin{eqnarray}
&\,& F_{T,L}^{CC,\,NC}(x,Q^2)  = \nonumber \\ &\,& 
  \frac{Q^2}{4\pi^2}  \int_0^1 dz \int d^2  \rr |\Psi^{W,Z}_{T,L}(\rr,z,Q^2)|^2 \sigma^{dp}(\rr,x)\,\,
\label{funcs}
\end{eqnarray} 
where $r$ denotes the transverse size of the dipole, $z$ is the longitudinal momentum fraction carried by a quark and  $\Psi^{W,Z}_{T,L}$ are proportional to the wave functions of the virtual charged or neutral gauge bosons corresponding to their transverse or longitudinal polarizations. Explicit expressions for $\Psi^{W,Z}_{T,L}$ are given, e.g., in Ref. \cite{nik_neu}.   Furthermore, $\sigma^{dp}$ describes the interaction of the  color dipole with the target and 
 encodes all the information about the hadronic scattering, and thus about the non-linear and quantum effects in the hadron wave function. As discussed in detail in our previous studies \cite{hepp,vicdie},  perturbative Quantum Chromodynamics (pQCD) predicts that the small-$x$ gluons in a hadron wave function should form a Color Glass Condensate (CGC) \cite{bal,cgc},  which  is characterized by the limitation on the maximum phase-space parton density that can be reached in the hadron
wave function (parton saturation), with the transition being
specified  by a typical scale, which is energy dependent and is
called saturation scale $Q_{\mathrm{sat}}$.  
 In Ref. \cite{hepp} we have estimated the neutrino - hadron cross section considering the current phenomenological saturation models for $\sigma^{dp}$ and compared with the predictions obtained using the solution of the Balitsky - Kovchegov equation \cite{bal,kovchegov}, which describes the CGC evolution in the  mean field approximation.  
In this letter, as our goal is to estimate the magnitude of the non-linear effects in the kinematical region probed by the IceCube Observatory, we will consider in our analysis  the  phenomenological saturation model
proposed in Ref. \cite{GBW}, denoted GBW hereafter, which encodes the main properties of the
saturation approaches. 
In the GBW model the dipole - hadron cross section is   parametrized
as follows
\begin{equation}\label{eq:gbw}
\sigma^{dp}_{GBW}(\rr,x)=\sigma_0\,[1-e^{-r^2Q_{\mathrm{sat}}^2(x)/4}],
\end{equation}
where the saturation scale is given by $Q_{\mathrm{sat}}^2=Q_0^2\left(x_0/x\right)^{\lambda}$,
$x_0$ is the value of the Bjorken $x$ in the beginning of the evolution and $\lambda$ is the
saturation exponent. The parameters $\sigma_0$, $x_0$ and $\lambda$ are obtained by fitting the $ep$ HERA data. In our study we assume the values obtained in Ref. \cite{Koslov07}, where the GBW model was updated in order to describe more recent data. The linear limit of the GBW model, which is obtained disregarding the non-linear effects, is given by 
\begin{equation}\label{eq:gbwlin}
\sigma^{dp}_{GBW_{lin}}(\rr,x)=\sigma_0\,\frac{r^2Q_{\mathrm{sat}}^2(x)}{4}\,\,.
\end{equation}
Although the linear limit of the GBW model is not able to describe the HERA data, 
a comparison of its predictions with the full model one is illuminating, since allows to directly quantify the contribution of the non-linear effects for a given observable. Another possibility to describe the  linear regime using the color dipole approach is to use that in the leading logarithmic approximation the dipole - hadron cross section  is  directly related to the target gluon distribution $xg$ as follows (See, e.g. Ref. \cite{Predazzi}): 
\begin{eqnarray}
\sigma_{dp}(\rr, x) =  {\frac{\pi^2}{3}} \rr^2 \alpha_s \, x g(x, 10/\rr^2) \,\,,
\label{gluongrv}
\end{eqnarray}
which satisfies the property known as color transparency, i.e. $\sigma_{dp}$ vanishes $\propto \rr^2$ at small separations. Such approximation is valid at small values of $x$ and large values of $Q^2$, which is the region probed in neutrino - hadron  interactions at high energies. If we assume that $xg$ is a solution of the DGLAP evolution equations, the use of this expression as input in our calculations implies 
that we are disregarding  non-linear QCD effects, associated to the high gluon density present at small-$x$ (large energies). In what follows we assume  that the   gluon distribution  is given by the CT10 parametrization \cite{ct10} and  that the resulting predictions correspond to the linear QCD dynamics, denoting them by Color Transp in the plots. The main advantage of this model in comparison to the GBW Lin one is that the gluon distribution used as input to calculate $\sigma_{dp}$ has been obtained in a global fit of the current experimental data. Finally, as in Ref. \cite{vicdie2} we include in our calculations the contributions associated to the heavy quarks, which contribute significantly at high energies.

\begin{figure}[t]
\vspace{0.5cm}
\includegraphics[scale=0.33]{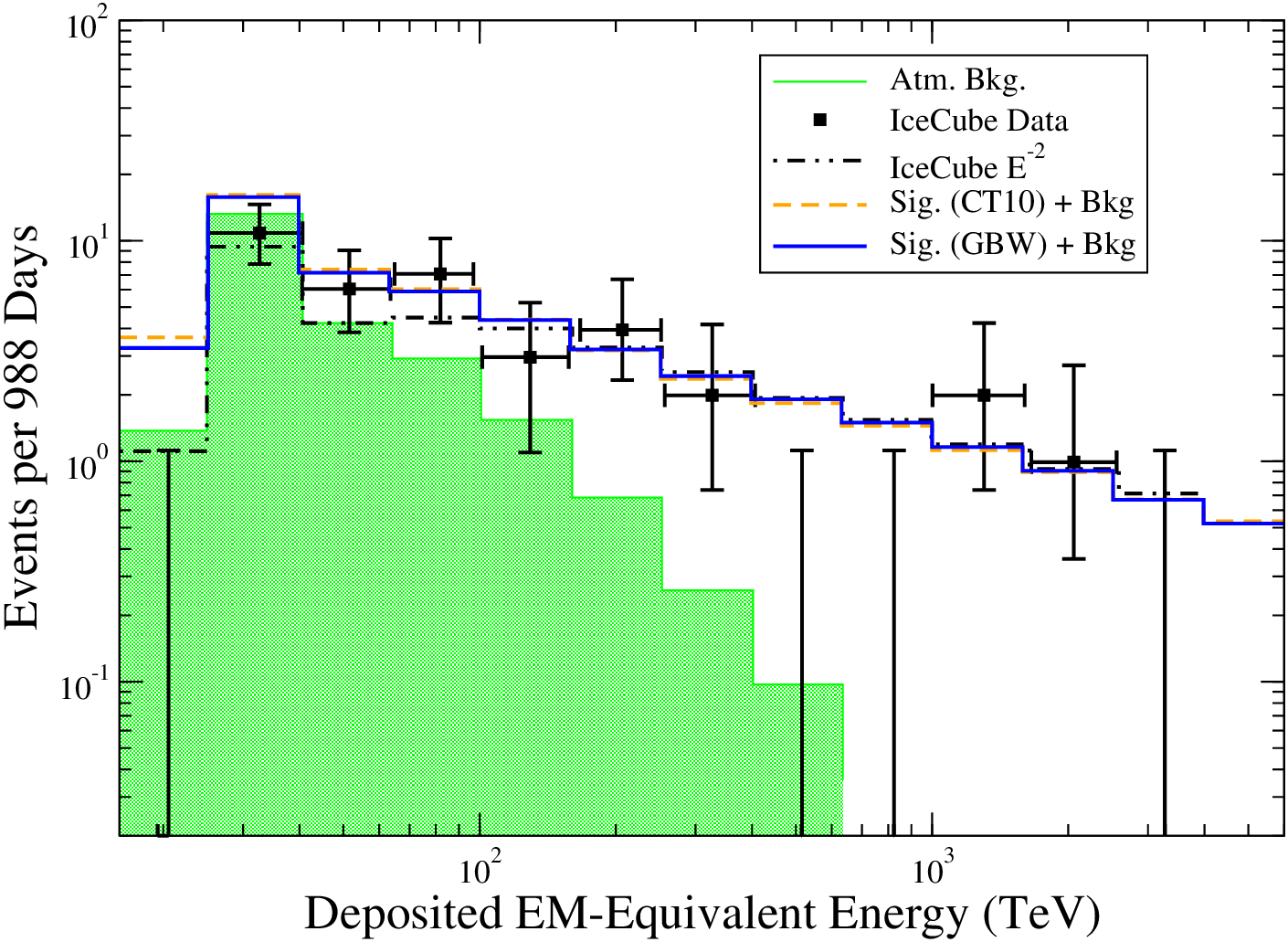}
\caption{(Color online) Comparison between the energy dependence of the number of UHE neutrinos events at IceCube predicted by the phenomenological non-linear GBW model (Solid blue line) and that obtained using the QCD improved parton model for the structure functions and the CT10 parametrization for the parton distribution functions (Dashed orange line). For comparison we also present { the $E^{-2}$ result  and the expected background reported by the IceCube Observatory}. IceCube data from Ref. \cite{icedata}. }
\label{fig:01}
\end{figure}

\begin{figure}[t]
\vspace{0.5cm}
\includegraphics[scale=0.33]{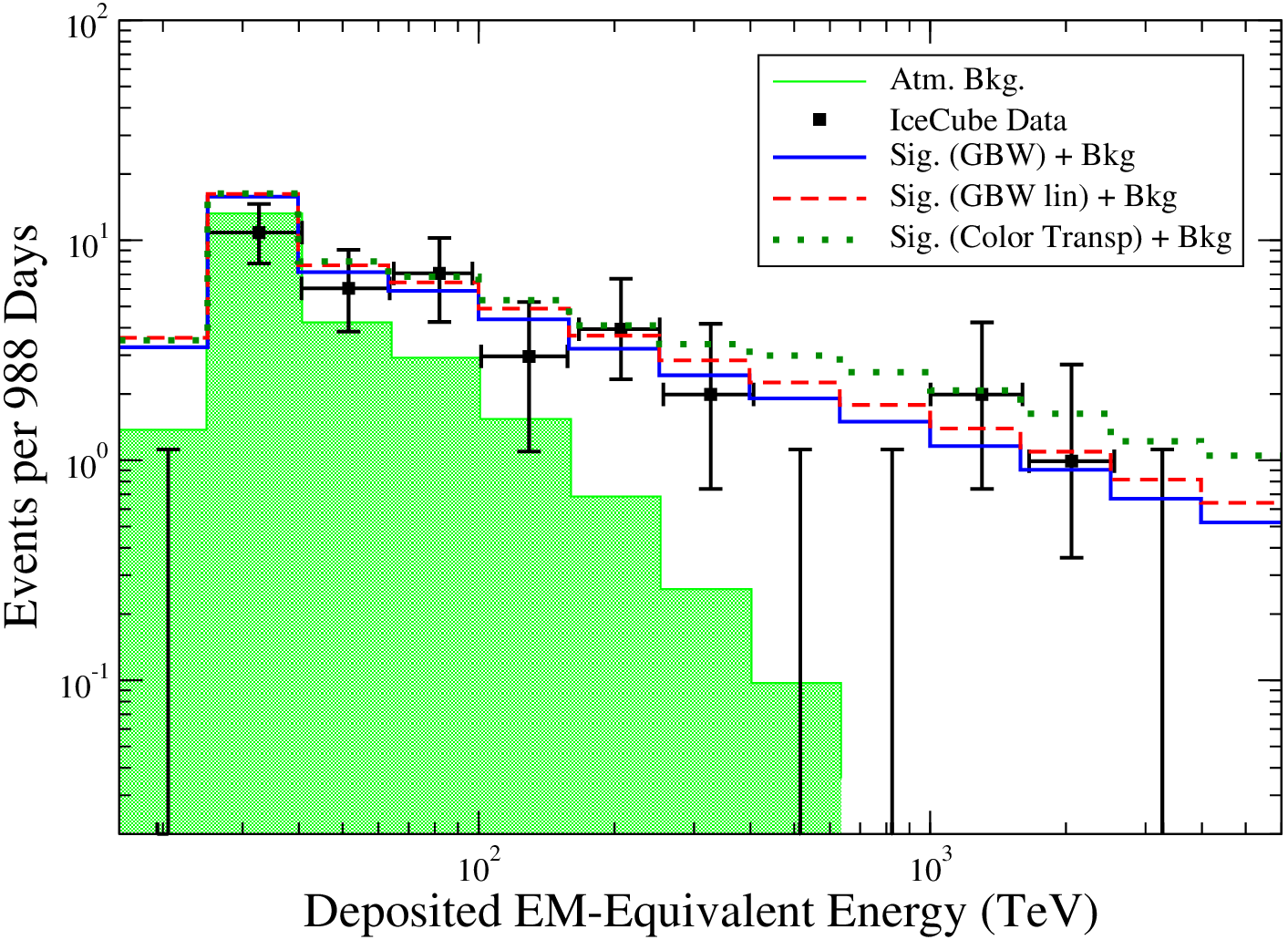}
\caption{(Color online) Comparison between the energy dependence of the number of UHE neutrinos events at IceCube predicted by the phenomenological non-linear GBW model (Solid blue line) and those obtained disregarding the non-linear QCD effects, denoted GBW lin (Dashed red line) and Color Transp (Dotted green line) in the figure. Green region is the expected background  reported by the  IceCube Observatory. IceCube data from Ref. \cite{icedata}.}
\label{fig:02}
\end{figure}

\begin{figure}[t]
\vspace{0.5cm}
\includegraphics[scale=0.33]{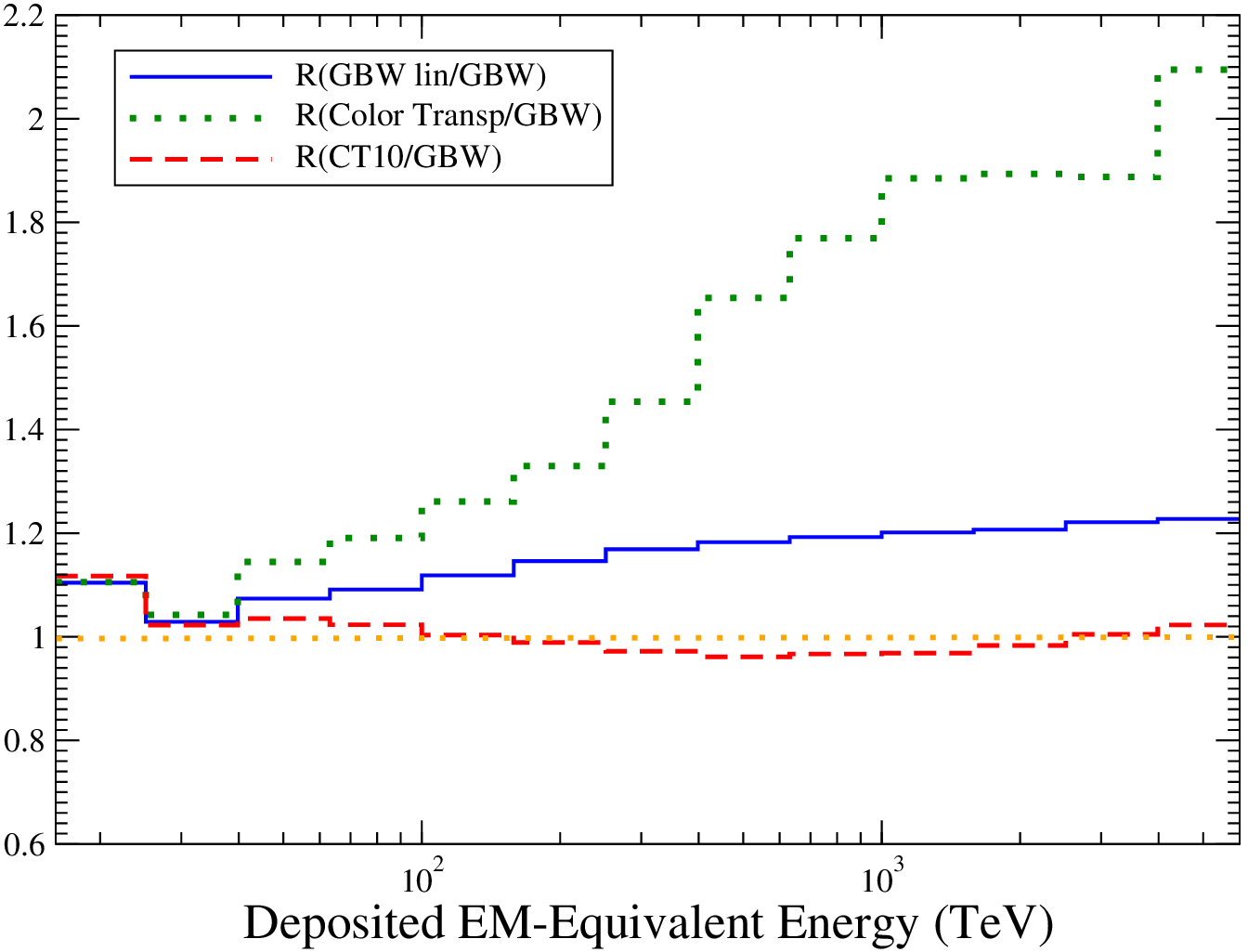}
\caption{(Color online) Energy dependence of the ratio between the  linear  and non-linear  predictions for the number of neutrino events at Icecube.}
\label{fig:03}
\end{figure}

In what follows we present our predictions for the number of neutrino events in the kinematical region probed by the IceCube Observatory. We estimate the neutrino - hadron cross section considering the models discussed above and use the results as input in the Eq.~(\ref{eq:dn}), which is integrated  with respect to the visible energy $E_{vis}$ deposited in IceCube detector. Considering the initial neutrino flux as described by Eq.~(\ref{eq:flux}), we must take into account   all the different possible neutrino reactions that produce signals inside IceCube due to this flux keeping in mind that the relation between  $E_{\nu}$ and $E_{vis}$ is different for each one of these processes. This is important since  the binning is done in the limits of integration in  $E_{vis}$. An explanation of this procedure  is presented in detail in Ref. \cite{Soni}, which we follow closely. In Fig.~ \ref{fig:01} we present our results for the dependence with $E_{vis}$ of the number of UHE neutrinos seen at IceCube as predicted by the integration of Eq.~(\ref{eq:dn}), added with the expected background as reported by IceCube Collaboration \cite{icedata}. For comparison we present the prediction obtained using the expression derived in the QCD improved parton model  for the structure functions and the CT10 parametrization \cite{ct10} for the parton distribution functions, which is similar to the results obtained in Ref. \cite{Soni} using the MSTW parametrization \cite{mstw}. Moreover, we also present the { $E^{-2}$ result  and the expected background reported by the IceCube Observatory}. We obtain that the GBW model describes quite well the experimental data, with its predictions being very similar to those obtained using the CT10 parametrization. In Fig. \ref{fig:02} we compare the GBW predictions with those obtained disregarding the non-linear QCD effects, denoted GBWlin and Color Transp in the figure. We obtain that these  predictions also are able to describe the data. In comparison to the GBW one, we have that all predictions are similar to low energies. In contrast, for higher energies, we obtain that the non-linear prediction is smaller than  the linear one. Both behaviours are expected theoretically, since the non-linear effects are predicted to contribute for high values of the parton densities, which should  be present at high energies. In order to quantify the contribution of the non-linear effects, {in Fig. \ref{fig:03} we present the energy dependence of the ratio between the linear ($N_{GBW_{lin}}$ and $N_{Color\,Transp}$) and non-linear ($N_{GBW}$) predictions for the number of neutrino events at Icecube. We present our predictions for the ratio as a function of the visible energy in the detector}. We obtain that the ratio increases with the visible energy $E_{vis}$, which is a direct consequence of the non-linear QCD effects present  in GBW model that limits the growth of neutrino-nucleon cross-section at high energies. The magnitude of the non-linear effects is strongly dependent on the model used to describe the linear regime of the QCD dynamics in the color dipole approach. In particular, if we consider the Color Transp model, we obtain that  the contribution of the non-linear effects is $\approx 80\%$ for $E_{vis}=1$ PeV, reaching 110 \% at $E_{vis}=6$ PeV. In contrast, for the GBW Lin model, the corresponding values are 20 \% and 22 \%, respectively.
These results indicate that the contribution of the non-linear QCD effects is not negligible in the kinematical region probed by the IceCube Observatory and should be considered as a source of theoretical uncertainty  in the  analysis of ultrahigh energy neutrino events. However, a final conclusion about the presence or not of the non-linear effects in the kinematical region probed by Icecube still is not possible. In particular, due to the large theoretical uncertainty present in the predictions for high energies obtained using the 
QCD improved parton model, associated to the uncertainties in the shape of light quark and gluon distributions in the small-$x$ and large-$Q^2$ regions. As demonstrated in Ref. \cite{Soni}, the uncertainty band in the predictions for number of neutrino events at Icecube associated to the 90 \% C. L. range of the parton distribution functions is large, with our GBW prediction being within the band.  
In particular, our GBW prediction is very similar to that obtained using the central  CT10 values for the parton distribution functions, as demonstrated in Fig. \ref{fig:03}, where we present the ratio between the corresponding predictions for the number of neutrino events as a function of the visible energy in the detector. Consequently, the discrimination between linear and non-linear QCD dynamics at Icecube through the analysis of the number of neutrino events will be a hard task, independently of the presence and magnitude of the non-linear effects.

Finally, lets summarize our results and conclusions. After the discovery of UHE diffuse astrophysical  neutrinos by the IceCube Observatory, the neutrino astrophysics lives now a new era, with the natural the next steps being the determination of the sources and possible production mechanism that would lead to such neutrino flux. However, in order to clarify precisely these aspects it is of utmost importance to determine the sources of uncertainties in the calculation of the number of ultrahigh neutrino events. In this letter we have estimated, for the first time, this quantity considering  non-linear QCD effects, which are expected to contribute for the QCD dynamics at high energies and, consequently, to modify the energy dependence of the neutrino - hadron cross section. We demonstrate that the phenomenological GBW model is able to describe the current IceCube data  and that the contribution of the non-linear effects is non-negligible  at visible energies larger than 2 PeV.

\section*{Acknowledgements} 
This work was  partially financed by the Brazilian funding
agencies CNPq, CAPES and FAPERGS.

\end{document}